\documentclass[10pt, conference, compsocconf]{IEEEtran}

\ifCLASSINFOpdf

\else

\fi

\usepackage{amsthm}
\usepackage{graphicx}  
\usepackage{subfigure}
\usepackage{booktabs}
\usepackage{multirow}
\usepackage{array}
\usepackage{changepage}
\usepackage{listings}
\usepackage{xcolor}
\usepackage{url}
\usepackage{setspace}
\usepackage{epstopdf}
\usepackage{amsmath, amssymb}
\usepackage{algorithm} 
\usepackage{algorithmic} 

\newcommand{\ie}{{\it i.e.,}}
\newcommand{\eg}{{\it e.g.,}}
\newcommand{\etc}{{\it etc}}

\begin{document}

\title{An Intelligent QoS Identification for Untrustworthy Web Services \\Via Two-phase Neural Networks}

\author{\IEEEauthorblockN{Weidong Wang$^{1}$, Liqiang Wang$^{2}$, Wei Lu$^{1}$}
\IEEEauthorblockA{$^{1}$School of Software Engineering, Beijing Jiaotong University, Beijing, China, 100044. \\ Email: \{wangweidong,luwei\}@bjtu.edu.cn\\
$^{2}$Department of Computer Science, University of Central Florida, Orlando, FL, USA, 32816. \\Email: lwang@cs.ucf.edu}}

\maketitle

\begin{abstract}

QoS identification for untrustworthy Web services is critical in QoS management in the service computing since the performance of untrustworthy Web services may result in QoS downgrade. The key issue is to intelligently learn the characteristics of trustworthy Web services from different QoS levels, then to identify the untrustworthy ones according to the characteristics of QoS metrics. As one of the intelligent identification approaches, deep neural network has emerged as a powerful technique in recent years. In this paper, we propose a novel two-phase neural network model to identify the untrustworthy Web services. In the first phase, Web services are collected from the published QoS dataset. Then, we design a feedforward neural network model to build the classifier for Web services with different QoS levels. In the second phase, we employ a probabilistic neural network (PNN) model to identify the untrustworthy Web services from each classification. The experimental results show the proposed approach has 90.5\% identification ratio far higher than other competing approaches.

\end{abstract}


\IEEEpeerreviewmaketitle

\maketitle

\IEEEdisplaynotcompsoctitleabstractindextext

\IEEEpeerreviewmaketitle
\section{Introduction}
\label{sec:intro}

Web services are self-contained, self-describing, modular applications, and reusable software components that are distributed and programmatically accessible over the Internet \cite{e1}. The success of Web service invocation largely depends on its QoS (Quality of Service) \cite{g1}. The QoS has become a significant factor in Web service management tasks since it indicates the critical features of a Web service such as reliability, throughput, and response time \cite{g3}. Indeed, the QoS-based Web service management tasks have been widely utilized to model and evaluate the non-functional features of a Web service \cite{g2}. In the service-oriented environment, there are often multiple functionally equivalent or similar Web services with different QoS from service providers or third party agents, which obviously span diverse organizations and computing platforms \cite{g4}. Service providers or third party agents may fail partially or fully in delivering the promised QoS at runtime \cite{e2}. On the other hand, it is not easy for users to identify the untrustworthy Web services \cite{g5}.

There are two reasons as follows. 1) The QoS information published by service providers may be unauthentic or misleading, or partially dependent on testing results in a particular period or a particular geographic area. 2) The QoS information of untrustworthy Web services identified by service users may be inaccurate, which should be primarily determined by domain experts.

In the QoS research, there are five main ways to obtain QoS of untrustworthy Web services, which are QoS collection \cite{g6,g7}, QoS monitoring \cite{g8,g9}, QoS prediction \cite{g10,g11}, QoS evaluation \cite{g12,g13}, and QoS management \cite{g14,g15}. Since the QoS collection approaches mainly focus on testing the quality of Web services under various environments, it may be confined by the high cost of testing environments and human resource. QoS monitoring approaches could become unrealistic for users to identify untrustworthy Web services because QoS may be unknown before Web services are executed. QoS prediction approaches rely on a great deal of historical data (\eg\ Web service location, invocation time, and environment \etc.) for accurately predicting untrustworthy Web services. QoS evaluation approaches could take huge human cost on identifying the untrustworthy Web services as these approaches entirely depend on investigating quality information from a great number of real-world Web services. QoS management approaches are more promising, which identify untrustworthy Web services by detecting inconsistency between delivered quality information and actual quality information.

Actually, in service computing, the QoS management approaches have been widely used in quality information detection and identification \cite{g14,g15,g16}. Most of previous researches employed statistic strategies and diagnostic strategies to identify inconsistency between the delivered QoS values and the expected QoS values of Web services. However, little work investigated how to identify untrustworthy Web services. If the untrustworthy Web services with QoS information cannot be accurately identified, any effective QoS management approach will become invalid since these untrustworthy Web services may result in QoS downgrade. Hence, an effective QoS-based identification approach for untrustworthy Web Services is very essential in QoS management process.

Complementary to previous QoS management approaches, which mainly focus on quality inconsistency or service-level agreement (SLA) violation detection, we propose a novel QoS-based approach to identify untrustworthy Web services via two-phase neural networks. The main idea is to learn the characteristics of trustworthy Web services from the trustworthy dataset by considering correlations of multiple QoS metrics. In the process, we build the trustworthy Web service dataset, which contains the typical Web services collection from public dataset \cite{g25}. Then, we propose a novel two-phase neural network model to identify the untrustworthy Web services. The contributions of this paper can be summarized as follows.


\begin{itemize}
  \item To classify QoS level of Web services, we design a novel feedforward neural network model to classify Web services. Compared with traditional models, our customized model typically considers correlations among the QoS metrics (\eg\ response time, availability, and throughput \etc.) according to the characteristics of trustworthy Web services.
    \item To identify untrustworthy Web services, we employ a probabilistic neural network (PNN) model to identify the untrustworthy Web services in each classification. Unlike traditional identification approaches, the PNN model is combined with multiple QoS metrics, and can ensure the accuracy.
\end{itemize}

The rest of this paper is organized as follows. Section \ref{sec2} describes the details of the proposed approach. Section \ref{sec3} describes the experiment results. Section \ref{sec4} describes the discussion of related work. Section \ref{sec5} concludes this paper and outlines the future work.



\section{Two-Phase Neural Networks}
\label{sec2}

\subsection{Overview}
\label{sec2:1}

\begin{figure}[t]
  \centering
  \includegraphics[width=8.5cm]{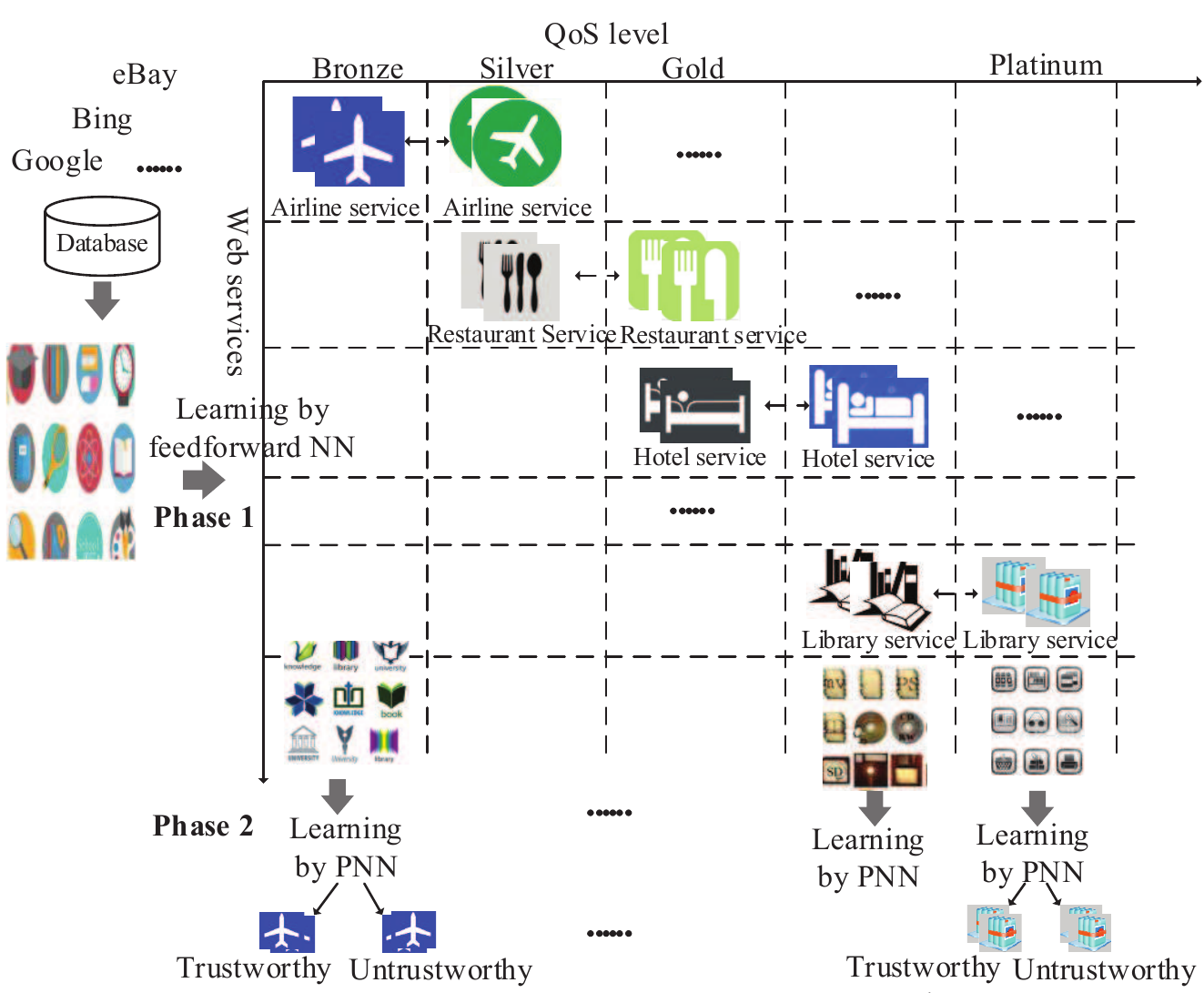}
  \caption{Framework of QoS-based identification of untrustworthy Web services.}\label{fig:1}
\end{figure}

This section introduces our two-phase neural networks. The framework of the proposed approach is shown in Figure \ref{fig:1}. In the first phase, the QoS values on each Web service from the Internet and databases such as the dataset \cite{g25} are classified based on their overall quality rating. Based on the work \cite{g22}, classification scheme associates each Web service to a particular service group (\eg\ bronze, silver, gold, platinum \etc.). We further extend the feedforward neural network as a classifier for Web services according to the characteristics of QoS metrics. Compared with the previous work \cite{g22}, the proposed approach can achieve more accurate classification results by considering the correlations among QoS metrics.

In the second phase, for each category, we employ the probabilistic neural network (PNN) to identify the untrustworthy Web services by considering the correlations among their QoS metrics. Compared with traditional reputation-based approaches, our approach explores the following capacities. 1) As the number of malicious feedback dynamically increases, our approach has high stable performance since it not only considers the correlations among QoS metrics but also depends on reputation. 2) The two-phase learning can avoid noises from dataset such as large number of malicious Web services because the first-phase learning for classification can guarantee accurate ranking and the second-phase can benefit from the previous phase and obtain the characteristics of Web services from different rankings. More details will be described in Section \ref{sec2:2}.

\subsection{First-phase Learning for Classification}
\label{sec2:2}
To classify Web services based on their QoS metrics, we employ the feedforward neural network as a classifier because of its advantage of processing multiple QoS metrics. Some details are shown as follows.

\begin{figure*}[t]
  \centering
  \includegraphics[width=12cm]{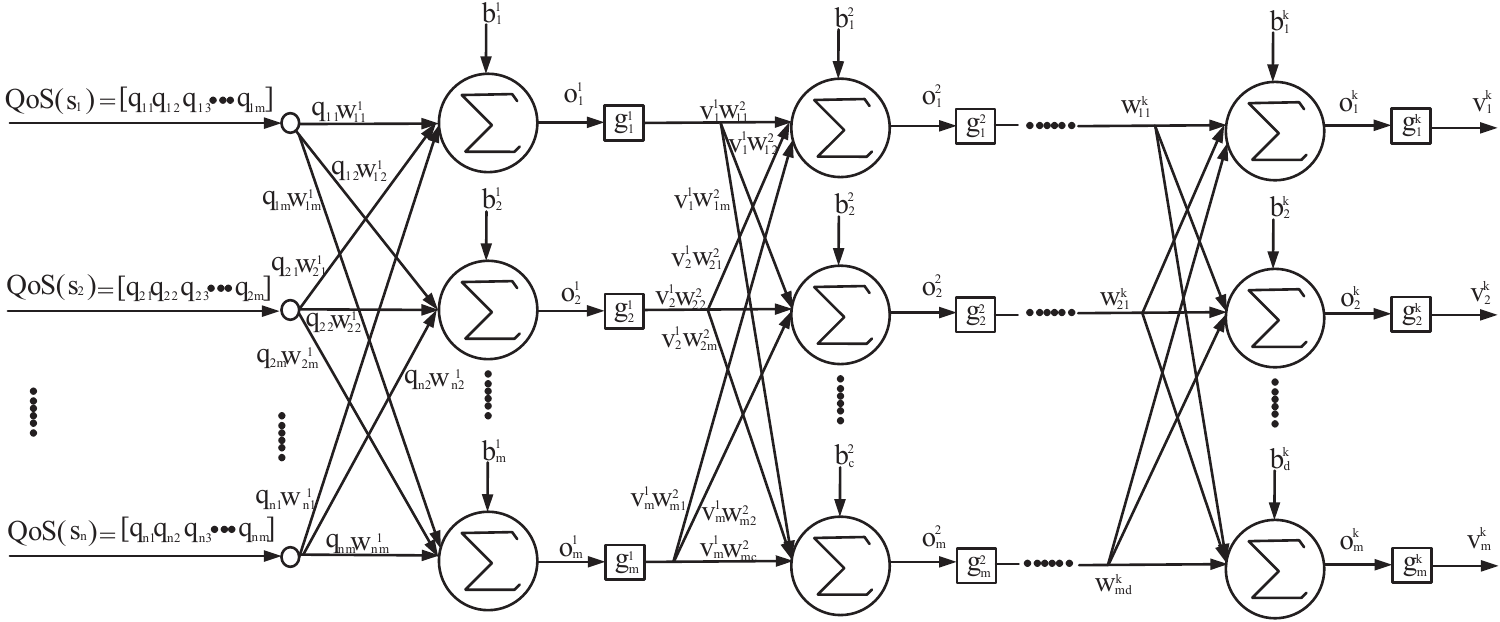}
  \caption{Learning framework for Phase-1.}\label{fig:2}
\end{figure*}

1) QoS normalization. A service class $S$ is a class of \emph{l} similar Web services \{$s_{1}$,$s_{2}$,$\cdots$,$s_{l}$\}, and a service $s_{i}$ may have \emph{m} QoS attributes (e.g. reputation, response time, price, etc.). The $j$-th QoS attribute of Web service $s_{i}$ is denoted as $q_{ij}$. An attribute is defined as a positive attribute, if a larger value means better service performance (e.g. reputation). An attribute is defined as a negative attribute, if a smaller value means better service performance (e.g. response time). Therefore, the attribute values of the services are normalized in the same class with the Simple Additive Weighting (SAW) method \cite{e3}, as shown in Equation (\ref{eq:1}):

\begin{equation}\label{eq:1}
\begin{cases}
\frac{q_{ij}-q_{min}}{q_{max}-q_{min}}  &  if\ q_{ij}\ is \ a \ positive\ attribute\\
\frac{q_{max}-q_{ij}}{q_{max}-q_{min}}  &  if\ q_{ij}\ is \ a \ negative\ attribute\\
1 & if\  q_{max}-q_{min}=0,
\end{cases}
\end{equation}

\noindent where $q_{max}$ and $q_{min}$ are the largest and smallest QoS attributes in the class.

2) First-phase learning framework. A feedforward neural network is a massive net consisting of a number of similar computing units, which are called service computation nodes. The morphology of a neural network can change the way how the nodes are interconnected and the operations performed on each node. Let $QoS(s_i)$ be the QoS metric vector, \ie\ $[q_{i1},q_{i2},\cdots,q_{ij},\cdots,q_{im}]$. As shown in Figure \ref{fig:2}, in an \emph{m}-layer feedforward neural network, all nodes in a layer are fully connected to the nodes in neighbor layers by weights, and adjustable parameters denote the strength of connections. The summation of weighted inputs to a node will be mapped by a nonlinear activation function $g_m^k$. There are no connections between nodes in the same layer. The QoS metric vector of each service is passed through the network in such a manner that the outputs of the nodes in the first layer become the inputs of the nodes in the second layer and so on. Mathematically, an \emph{m}-layer feedforward neural network can be expressed as follows,

\begin{equation}
\footnotesize
\begin{cases}
o^k=w^kv^{k-1}+b^k  \\
v^k=g^k(o^k) & (k=1,\cdots,m),
\end{cases}
\label{formu:1}
\end{equation}

\noindent where $v^0=QoS(s_i)=[q_{i1},q_{i2},\cdots,q_{im}]^T$ is the input vector; $o^k=[o_1^k,\cdots,o^k_{s_k}]$, $g^k=[g_1^k,\cdots,g^k_{s_k}]^T$, and $v^k=[a_1^k,\cdots,a^k_{s_k}]^T$ are the linear output vector of the summation, the activation function vector, and the output vector in the $k^{th}$ layer, respectively; $s_k$ is the number of nodes in the $k^{th}$ layer; $w^k$ and $b^k$ represent the weight matrix and the bias vector in the $k^{th}$ layer, which can be expressed as follows.

\begin{equation}
\footnotesize
\omega^k = \left(
  \begin{array}{ccc}
          \omega_{11}^k & \cdots & \omega_{1s_{k-1}}^k\\
          \vdots & \ddots & \vdots\\
          \omega_{s_k1}^k & \cdots & \omega^k_{s_ks_{k-1}}
  \end{array}
  \right)\ \textbf{and}\ b^k = \left(
  \begin{array}{c}
          b_1^k  \\
          \vdots \\
          b_{s_{k}}^k  \\
  \end{array}
  \right),
\label{formu7}
\end{equation}

\noindent where the $j^{th}$ row of $\omega^k$ is defined by $\omega^k_j=[\omega^k_{j1}, \omega^k_{j2}, \cdots, \omega^k_{js_{k-1}}]$. Therefore, the goal is to obtain these weights by training samples.

3) Training method. The well-known training method is the perturbation weights method, which randomly perturbs one weight and observes whether it enhances performance. However, the method may face some limitations as follows. 1) The reinforcement learning could be inefficient since the value change of one weight may contribute little to the overall performance in a large number of dimensions of QoS. 2) To acquire the improvement of performance, the large weight perturbations could lead to the worse output because the perturbations may be deviated from the right relative multiple weights for each attribute of QoS. Therefore, we employ the backpropagation method \cite{e7} customized for the weights acquisition on each QoS attribute. In this way, we can efficiently obtain better performance than the perturbing method as shown in the experiments of Section \ref{sec3}.

\subsection{Second-Phase Learning for Identification}
\label{sec3.3}

Based on the first-phase learning, we obtain the classification of Web services. For Web services in each classification, we need to identify the untrustworthy ones.

\begin{figure*}[htpb]
  \centering
  \includegraphics[width=12cm]{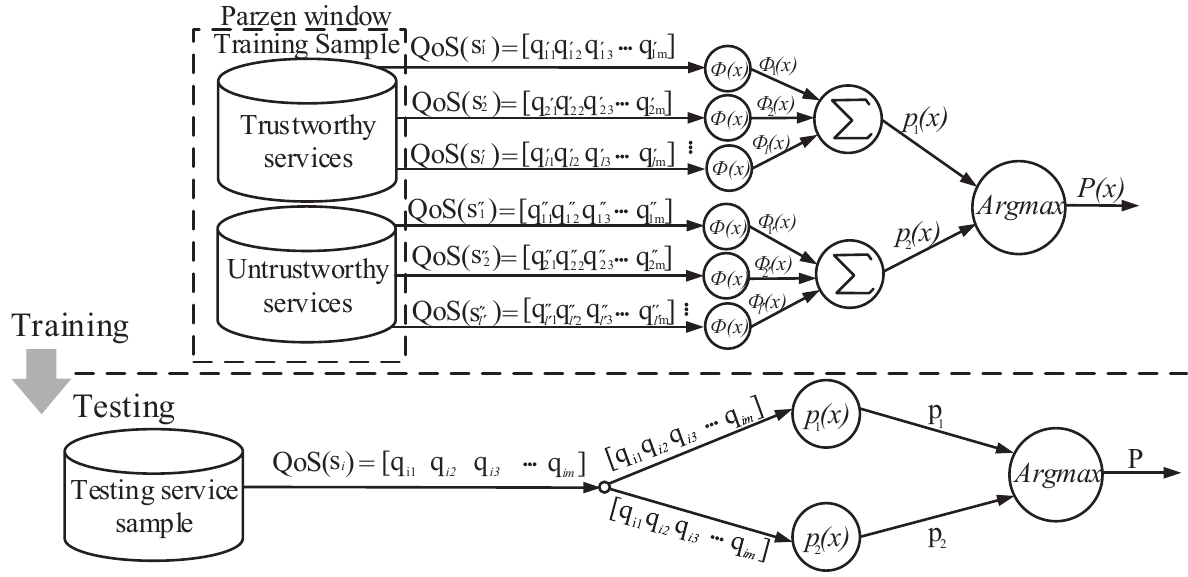}
  \caption{Learning framework for Phase-2.}\label{fig:3}
\end{figure*}

1) Second-phase learning framework. To identify the untrustworthy Web services, we employ the probabilistic neural network (PNN), which is based on the theory of Bayesian classification and the estimation of probability density function. The PNN consists of several sub-networks, each of which can be viewed as a parzen window estimator for the training samples. Actually, for each sample classified by the first neural network, we collect the labelled trustworthy Web services and untrustworthy Web services as the training sample of the second neural network. According to the Gaussian window function, the parzen window identifier makes a decision after calculating the probability density function of each Web service using the given training samples. The multi-category identifier decision is expressed as follows:

\begin{equation}
\footnotesize
\phi(s_i)= \frac{1}{(2\pi)^{1/2}\sigma}exp(-\frac{\|QoS_{c,j}-QoS(s_i)\|^2}{2\sigma^2}),
\label{formu:1}
\end{equation}

\noindent where $QoS_{c,j}$ denotes the given QoS metric vector of the $j$-th Web service of the $c$-th category. $\sigma$ is the smoothing factor. The summation layer neurons compute the maximum likelihood of category $c$. According to the above probability, $s_i$ can be classified by summarizing and averaging the outputs of all neurons that belong to the same category.

\begin{equation}
\footnotesize
p_c(s_i)= \frac{1}{(2\pi)^{1/2}\sigma}\frac{1}{n_c}\sum_{i=1}^{n_c}exp(-\frac{\|QoS_{c,j}-QoS(s_i)\|^2}{2\sigma^2}),
\label{formu:1}
\end{equation}

\noindent where $p_c(s_i)$ denotes the probability that the Web service $s_i$ belongs to the $c$-th category. $n_{c}$ denotes the total number of samples in category $c$. The input instance with unknown category is propagated to the pattern layer. Once each node in the pattern layer receives the input, the output of the node will be computed by Equation (\ref{formu6}) as follows.

\begin{equation}
\footnotesize
p(s_i)=argmax\{p_c(s_i)\}, c \in \{1,2\}
\label{formu6}
\end{equation}

\noindent where $p(s_i)$ denotes the probability, which takes the maximum between $p_1(s_i)$ and $p_2(s_i)$. Note that $c=1$ denotes the untrustworthy category and $c=2$ denotes the trustworthy category.

2) Training method. There is neither iteration nor computation of weights. For a large number of $Gaussians$ in a sum, the error accumulation may be significant. Thus the feature vectors in each category may be reduced by making $\sigma$ larger. However, due to each pattern layer Gaussian window density function $p_c(s_i)$ being derived from a group of training samples, the PNN is limited to applications involving relatively small datasets. Large datasets may lead to large and complex network, which would result in adverse impact on computational complexity. In addition, this could saturate the feature space with overlapping Gaussian function that may increase the rate of misclassification. Meanwhile, for the corner case such as $p_1(s_i) = p_2(s_i)$, if $p_1(s_i) = p_2(s_i) \geq 0.5$, we need force to determine the $s_i$ belongs to the untrustworthy category, or vice versus.

\section{Experiments}
\label{sec3}
In the section, we conducted an extensive experimental evaluation on the proposed approach for untrustworthy Web service identification. All the learning algorithms from the above two phases are implemented by MATLAB R2015B and the experiments run on HP 8280 with four Intel Cores i5-2400 of 3.1GHz and with 8GB RAM.

The experiments mainly consist of two parts: 1) the proposed neural network method is compared with other popular competitors; 2) The impacts of different parameters to the identification accuracy are evaluated.

\subsection{Experiment Setup}
\label{sec3.1}

\begin{figure*}[th]
\centering
\subfigure[100\% malicious feedbacks]{\includegraphics[width=5.7cm]{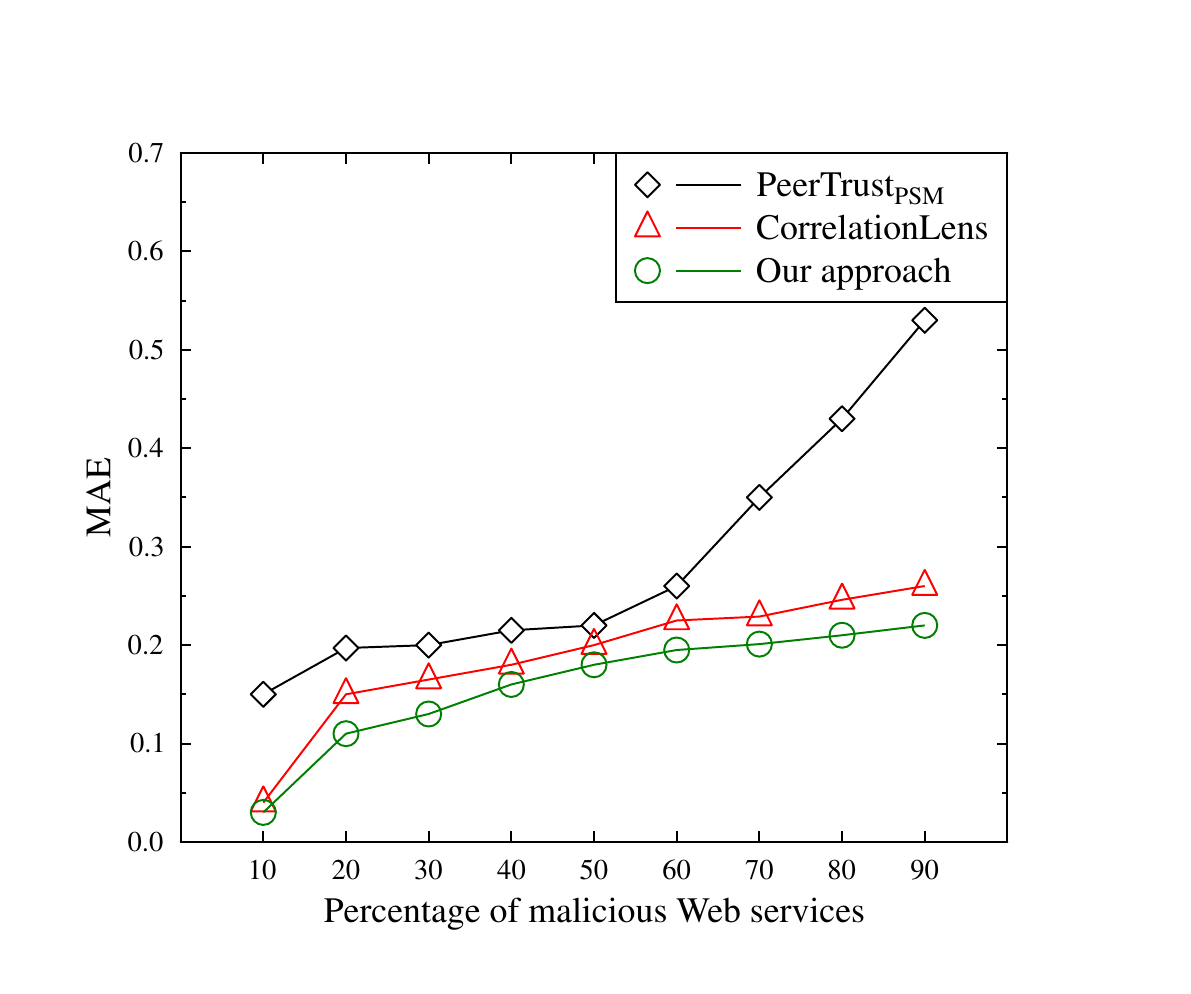}}
\subfigure[10\% malicious Web services]{\includegraphics[width=5.7cm]{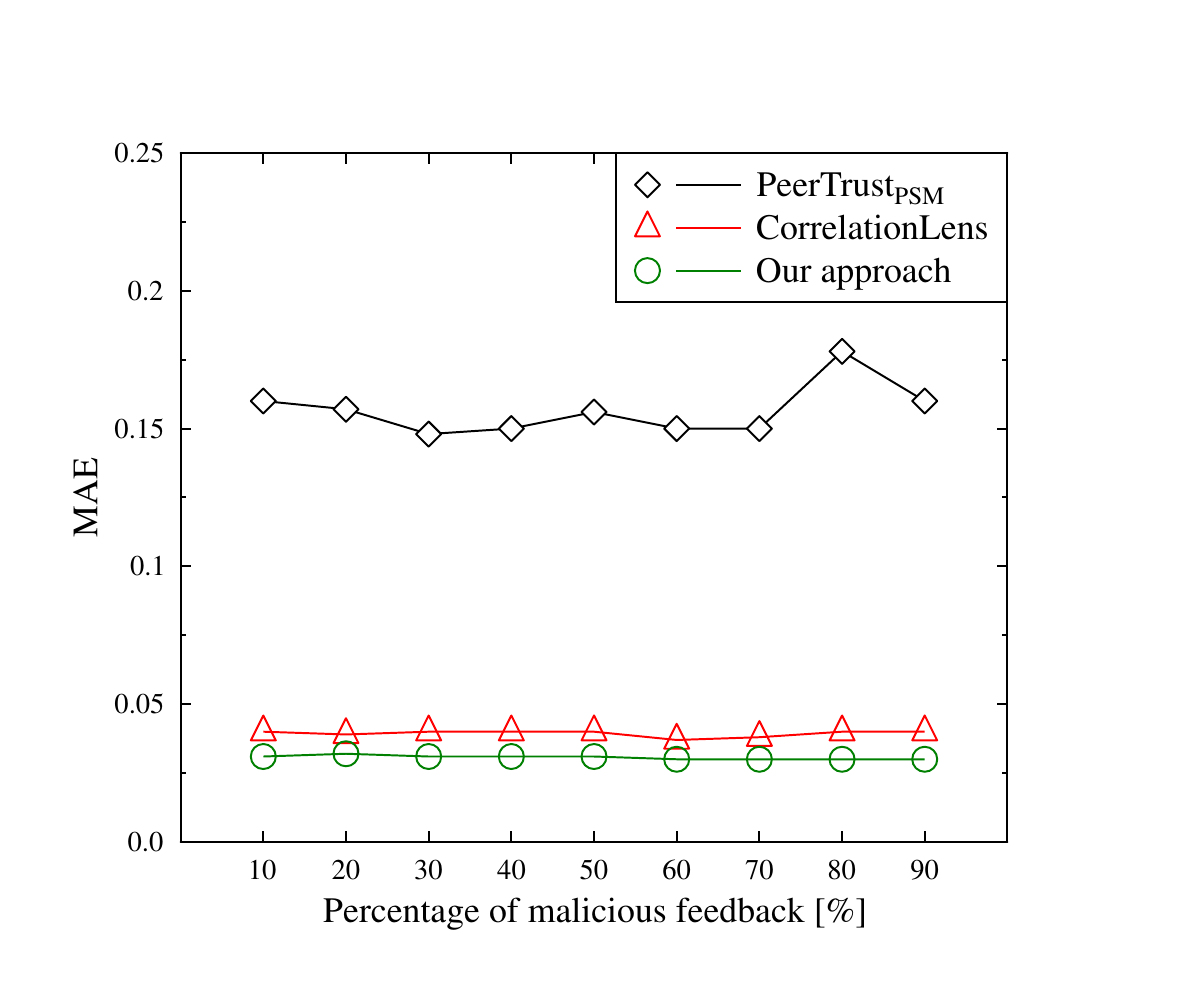}}
\subfigure[30\% malicious Web services]{\includegraphics[width=5.7cm]{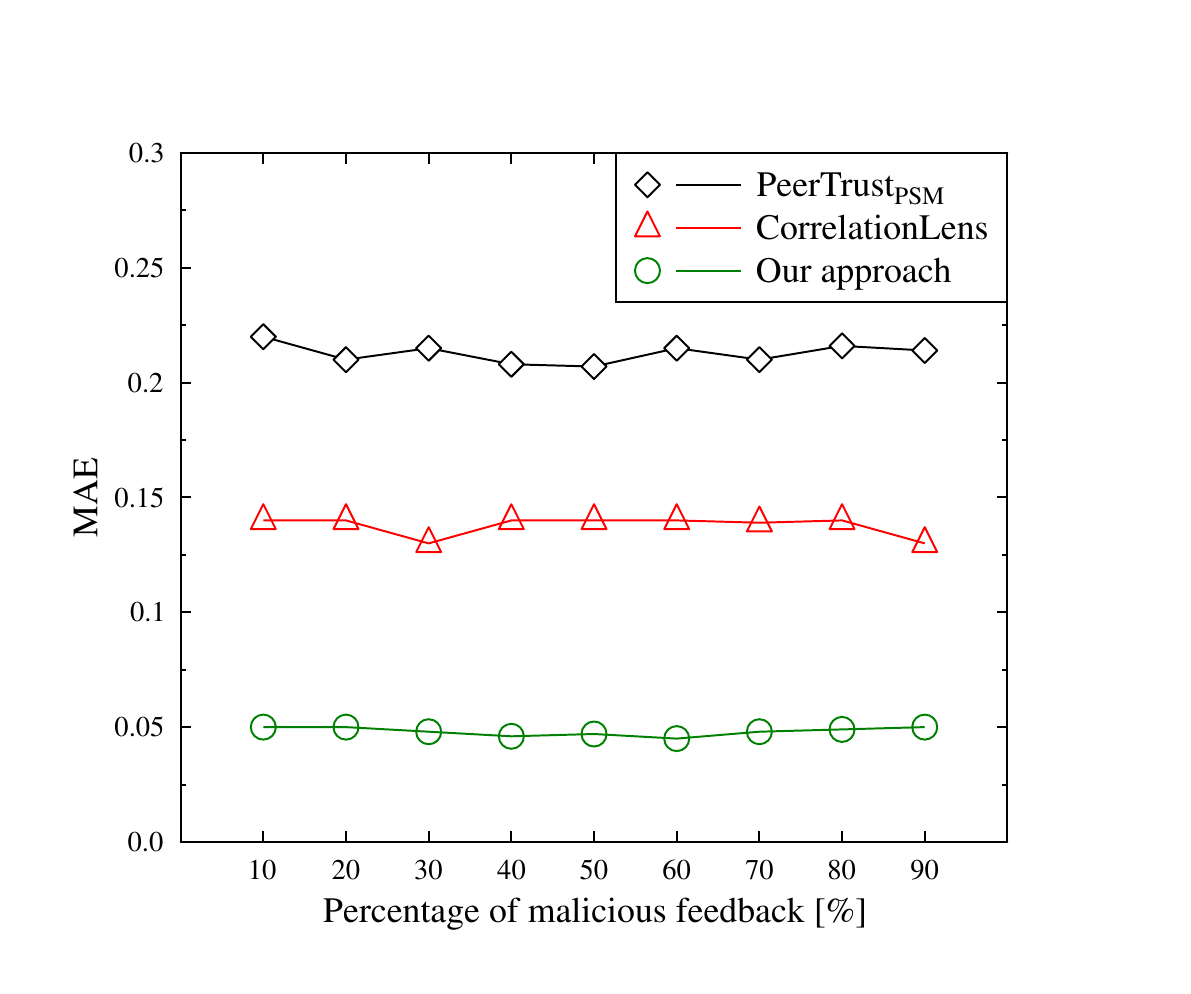}}
\subfigure[50\% malicious Web services]{\includegraphics[width=5.7cm]{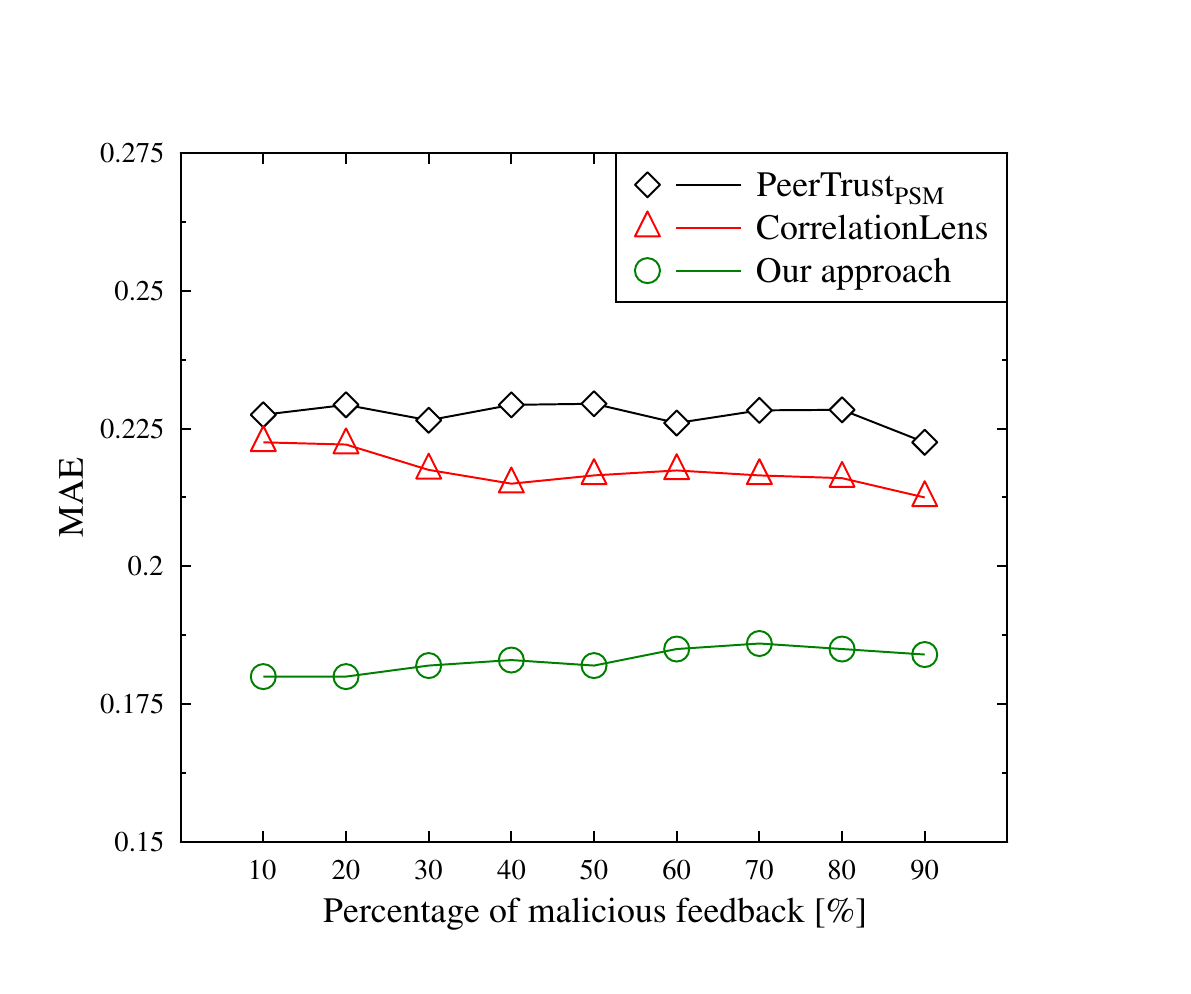}}
\subfigure[70\% malicious Web services]{\includegraphics[width=5.7cm]{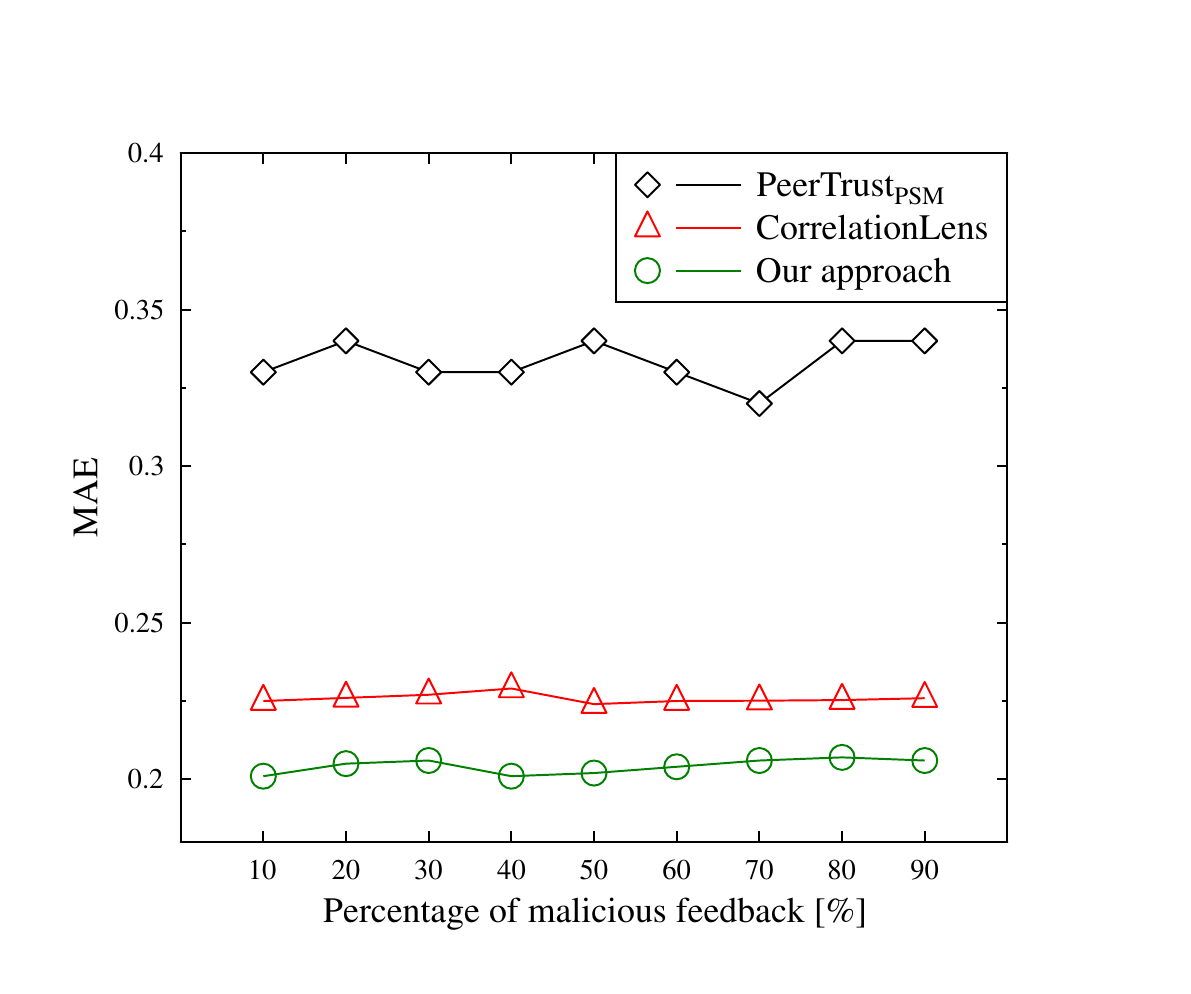}}
\subfigure[90\% malicious Web services]{\includegraphics[width=5.7cm]{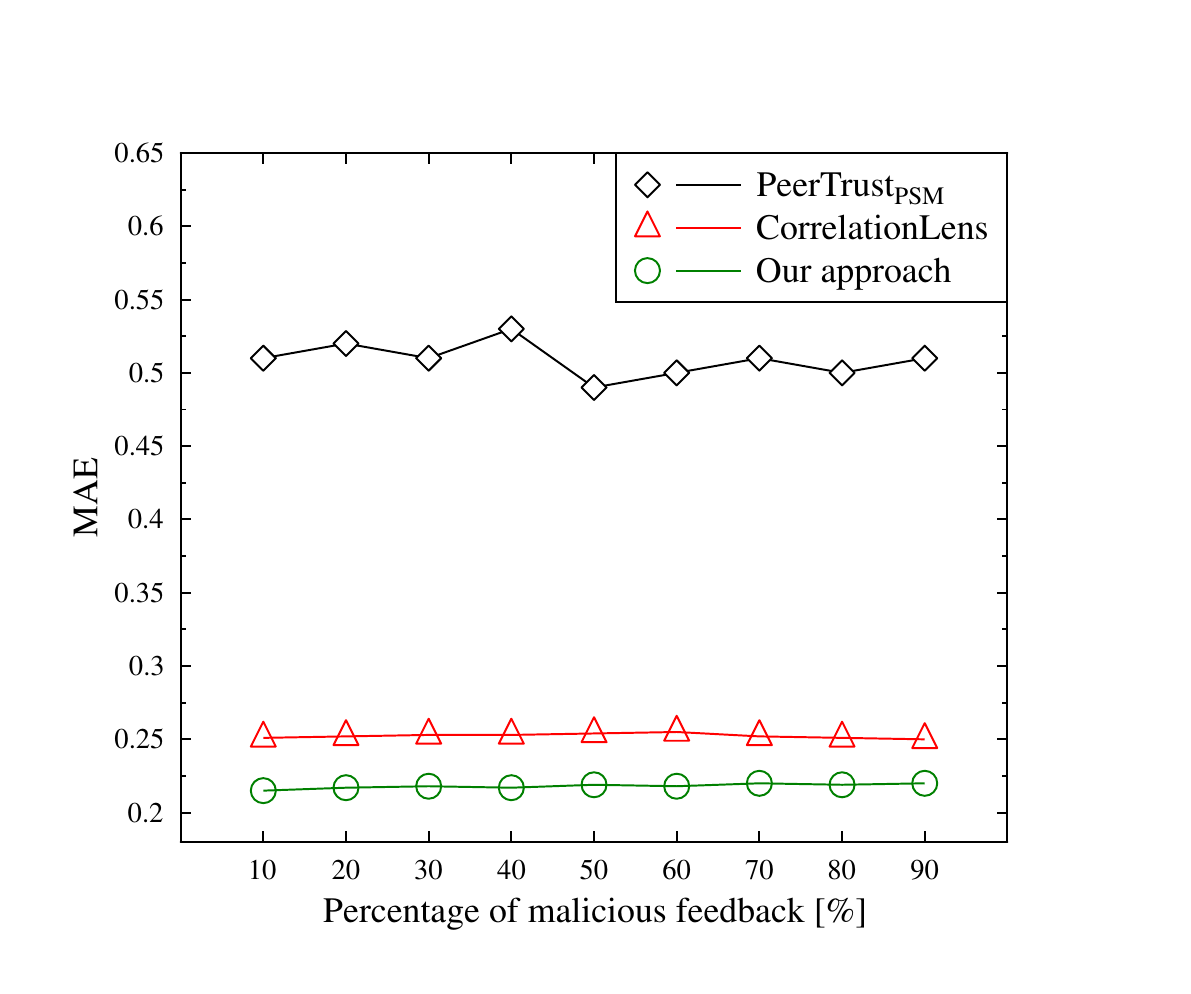}}
\caption{Comparisons with varied ratio of malicious Web services and reputation feedback (a smaller MAE value means a better performance).} \label{fig:4}
\end{figure*}

1) Data source. To evaluate the identification capability of the proposed approach, we created a controlled environment containing Web services that mainly derive from two aspects as follows: a) the well-known companies such as Amazon and b) typical Web service datasets (\eg\ QWS Dataset, \url{http://www.uoguelph.ca/~qmahmoud/qws/}). To obtain QoS values of Web services, we directly access the Web portal of the QWS Dataset, which comprises measurements of 9 QoS attributes for 2507 typical real-world web services for web service researchers.

2) Evaluation metric. We apply identification ratio to measure the accuracy of the proposed approach, which can be calculated by using the number of Web services with correct identification. Identification ratio is defined as follows.

\begin{equation}
\footnotesize
Identification\ ratio = \frac{N_{identified}}{N}, \\
\label{formu24}
\end{equation}

\noindent where $N_{identified}$ denotes the total number of Web services with correct identification, and $N$ denotes the total number of Web services. In addition, identification approaches must identify the untrustworthy Web service as accurately as possible. Therefore, differences between the identified Web services and the true performance of Web services are usually employed to evaluate identification accuracy. Mean Absolute Error (MAE) is widely adopted as an evaluation metric for the proposed approach. MAE is defined as follows.

\begin{equation}
\footnotesize
MAE = \frac{\sum_j|I_j-\hat{I_j}|}{J}, \\
\label{formu24}
\end{equation}

\noindent where $I_j$ denotes the identification ratio of untrustworthy Web services from the $j$-th testing sample. $\hat{I_j}$ denotes the ideal identification ratio of untrustworthy Web services obtained by users. $J$ is the number of testing samples.

3) Parameter configuration. In our experiments, the main parameters are shown as follows. a) Acceptable error value is used to determine whether the network is able to converge into a possible result or not. b) Learning rate is adopted to determine the speed of training process. c) The number of epochs is used to determine the possible accuracy. d) In the second learning phase, $\sigma$ is a smoothing factor, which may impact the performance of identification. The above parameters have to be fixed prior to the training process. Therefore, we may try different parameter values according to users' requirements.

\subsection{Comparisons}
\label{sec3.3}

\begin{table*}[t]
\centering
\caption{Identification ratio with different number of malicious Web services.}
\begin{tabular}{p{3.5cm}p{1cm}p{1cm}p{1cm}p{1cm}p{1cm}p{1cm}p{1cm}p{1.5cm}}\\
\toprule
Number of malicious services    & 10\%	 &20\%      &30\%       &40\%          &50\%           &60\%                 &Average           &Standard deviation       \\
\midrule
$PeerTrust_{PSM}$               &0.96	 &0.85     &0.76	     &0.60	  &0.58	      &0.56	             &71.83\%	              &16.47\%               \\
\emph{CorrelationLens}	        &0.95	 &0.87     &0.84	     &0.82	  &0.79	      &0.76	             &83.83\%	              &6.67\%	                   \\
\emph{Our approach}             &0.96	 &0.94     &0.92	     &0.91	  &0.87	      &0.83	             &90.50\%	              &4.76\%	                   \\
\bottomrule
\end{tabular}
\label{tab:1}
\end{table*}

To study the identification performance, we compare the proposed approach with two other identification approaches, $PeerTrust_{PSM}$ \cite{e4} and $CorrelationLens$ \cite{g18}.



\begin{itemize}
  \item \emph{$PeerTrust_{PSM}$ approach.} This approach is based on a personalized similarity measure (PSM) \cite{e4}. In the approach, the credibility of peer is based on the similarity between the feedback to all other peers.
  \item \emph{$CorrelationLens$ approach.} This approach is based on the probability theory to estimate the trustworthiness of Web services by leveraging the correlation information among various QoS metrics \cite{g18}.
  \item \emph{\emph{Our approach.}} The proposed approach employs two-phase neural network model to identify the untrustworthy Web services.
\end{itemize}

In this experiment, we fixed 5 hidden layers and 9 neuron nodes that received 9 QoS metrics of Web services from the training sample. Meanwhile, we used the sigmoid activation function for training. Figure \ref{fig:4}(a) shows the MAE of the reputation based on feedback from 2507 Web services with varied ratio of malicious ones. It is obvious that the proposed approach is significantly better than the $PeerTrust_{PSM}$ when the ratio of malicious Web services is more than 60\%. The $peerTrust_{PSM}$ may be confined on the number of malicious feedback since it depends on similarity computation with benign Web services. Moreover, the proposed approach is slightly better than the \emph{CorrelationLens} approach as the number of malicious feedbacks increases since the \emph{CorrelationLens} approach more or less relies on the hypothesis of distributions. However, our approach benefits from learning on real-world Web services for better performance. Figure \ref{fig:4}(b) to Figure \ref{fig:4}(f) show the MAE performance when the ratio of malicious Web services are fixed at 10\%, 30\%, 50\%, 70\%, 90\%, respectively. A noteworthy observation is that the proposed approach outperforms other approaches with the increasing number of malicious Web services. For example, in Figure \ref{fig:4}(c), the ratio of malicious Web services is fixed at 30\% and the number of percentage of malicious feedbacks varies from 10\% to 90\%, the MAE of the proposed approach is 0.05 far less than \emph{CorrelationLens} (0.14) and $PeerTrust_{PSM}$ (0.22). In addition, the proposed approach achieves a stable performance than the others in large number of malicious feedbacks and Web services. The proposed approach achieves good performance because of the identification via its two-phase learning process.

Table \ref{tab:1} shows that the maximal and minimal identification ratio of the proposed approach are 0.83 and 0.96, respectively, while the maximal and minimal identification ratio of the \emph{CorrelationLens} are 0.76 and 0.95. The average identification ratio of our approach is 90.50\%, and significantly higher than 83.83\% of the \emph{CorrelationLens} and 71.83\% of the $PeerTrust_{PSM}$. The standard deviation of our approach (4.76\%) is slightly smaller than the \emph{CorrelationLens} (6.67\%) and far smaller than the $PeerTrust_{PSM}$ (16.47\%). It means that the proposed approach is more accurate and stable than the \emph{CorrelationLens} and the $PeerTrust_{PSM}$.

\subsection{Impact of parameters}
\label{sec3.3}

\begin{figure}[th]
\centering
\subfigure[Impact of learning rate]{\includegraphics[width=4.2cm]{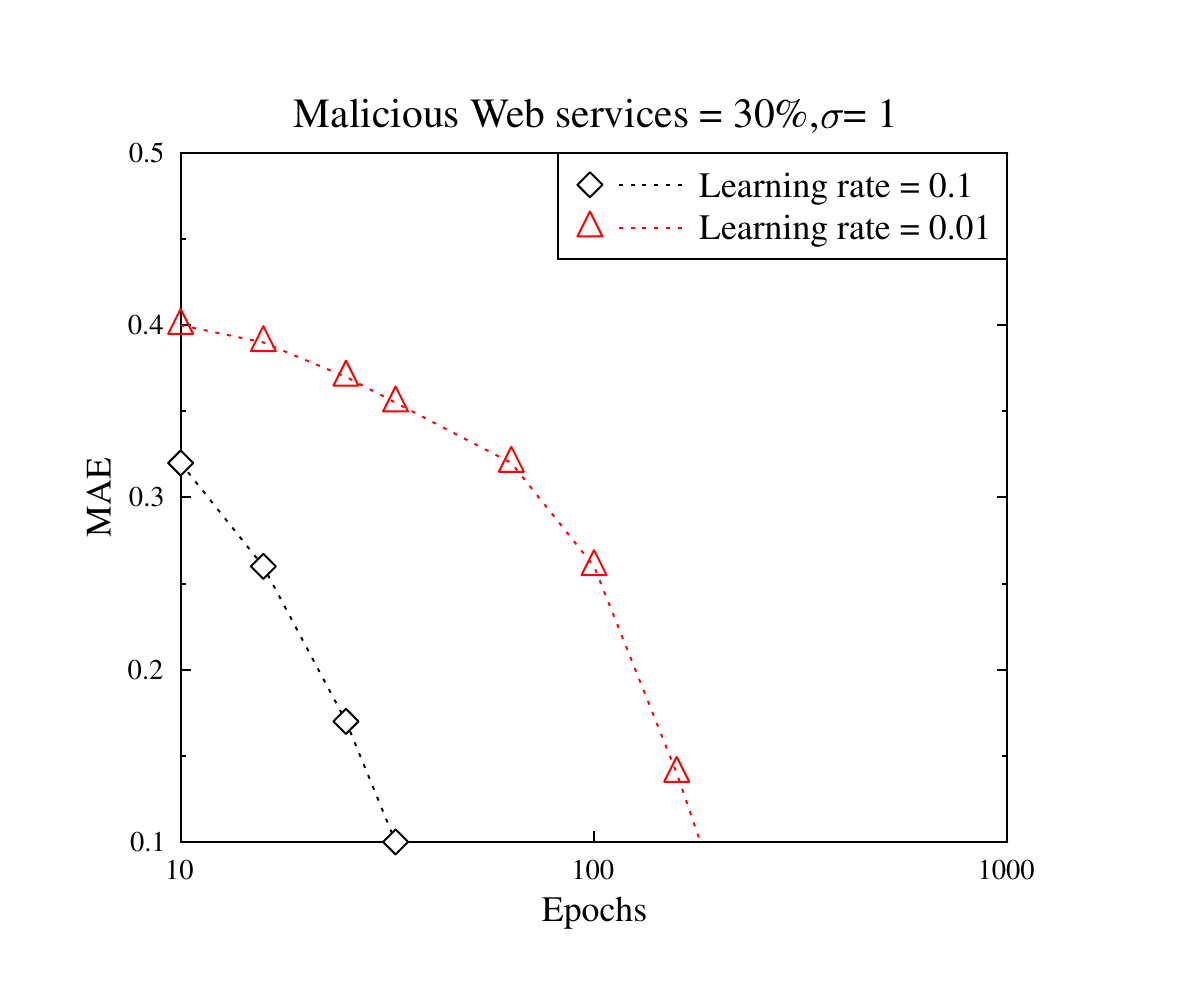}}
\subfigure[Impact of $\sigma$]{\includegraphics[width=4.2cm]{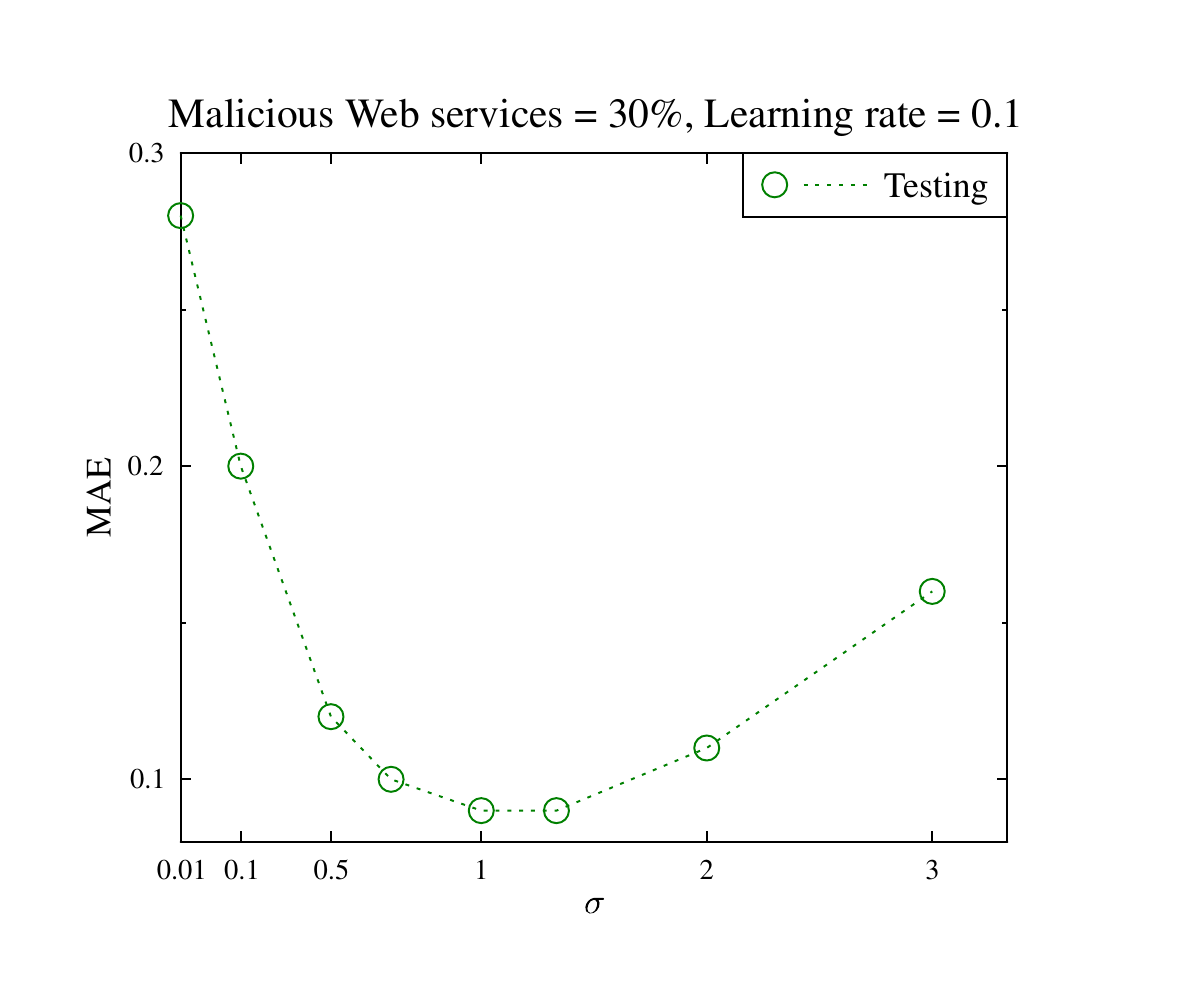}}
\caption{Impact of parameters.} \label{fig:5}
\end{figure}

Figure \ref{fig:5}(a) shows that learning rate can significantly impact on the processing time of neural network. For example, while obtaining the MAE = 0.10, the case where the learning rate equals 0.01 takes 251 iterations more than the case (49 iterations) where the learning rate equals 0.1. These results also show that the smaller the acceptable MAE is, the longer time the network will take. Therefore, while choosing the neural network as an acceptable solution for identifying the untrustworthy Web services, a trade-off between the learning rate and the acceptable accuracy should be considered since neither the fast speed and low accuracy nor the low speed and high accuracy will be accepted while meeting the users' requirements. In the above experiments, we try different combinations of both and finally choose the learning rate = 0.01 and the MAE approximately equals 0.1.

Figure \ref{fig:5}(b) shows that the value of smoothing factor $\sigma$ is significant since it determines whether the classification by the PNN is correct or not. The learning rate and the ratio of malicious Web services are fixed at 0.1 and 30\%, respectively. It is obvious that the MAE value is continuously decreasing as the value of $\sigma$ increases but it is not beyond 1 because the smaller value of $\sigma$ means the smaller window size of PNN and it may lead to overfitting. Moreover, as the value of $\sigma$ increases beyond 1, the MAE value may increase since the larger window size of PNN makes some trustworthy Web services to stand the untrustworthy side and it may result in incorrect classification. Therefore, in our experiment, the value of $\sigma$ is set to 1 for the purpose of high accuracy.

\section{Related Work and Discussion}
\label{sec4}

\subsection{QoS Management}
\label{sec:4.1}

The dynamic e-business vision calls for high QoS of Web services over the Internet. Delivering QoS on the Internet is a significant and critical challenge due to its dynamic and unpredictable nature. Unresolved QoS issues such as untrustworthy Web services with malicious QoS may cause critical transactional applications to suffer from unacceptable levels of performance degradation \cite{g26}. To address this problem, various QoS management approaches such as \cite{g14,g27} were proposed to manage trustworthy Web services. However, most of them are based on detecting the inconsistency between the delivered QoS information and practical QoS information. These approaches employ evaluating or diagnosing trustworthiness of Web services based on runtime-diagnosis or post-processing yet not previous trustworthiness identification. In addition, these approaches computing the trustworthiness of Web services are based on one QoS metric or various combinations of QoS metrics but not referring the correlations among the quality metrics, which may lead to somehow inaccurate results.

Different from the above approaches, we build the trustworthy Web service database by totally considering the correlations among the QoS metrics, then learn the characteristics of trustworthy Web services by feedforward neural network for the previous classification for enhancing the accuracy of the identification.

\subsection{Trustworthy Web Service Identification}
\label{sec:4.2}

In the field of service computing, the trust and reputation system becomes popular and may be more concerned by some research institutes due to its important functions that can assist the interactions and select trustworthy Web services among different parties. Hang et. al \cite{g17} utilized the beta-mixture distribution to model the quality of agent-based services for justifying whether a agent is trustworthy or not. The model may be confined on processing only one QoS metric at each time when interacting with other agents. While handing multiple QoS metrics, it would be time consuming. To enhance the performance under multiple QoS metrics, Nguyen et al. \cite{g19} proposed a Bayesian probability trust and reputation model to compute the trustworthiness of Web services by considering various combinations of QoS metrics, yet the model lacks of the correlation information that exists among the different QoS metrics. Such information may lead to over-estimated confidence in the trustworthy services. To address the problem, Mehdi et al. \cite{g18} used the multi-nomial distribution to define the number of pairs of QoS metrics for computing the trustworthiness by a hybrid dirichlet distribution, which can be used to handle the correlated QoS metrics to avoid over or under estimating the confidence of obtained trustworthy services.

Different from the above work, we identify the untrustworthy Web services by totally considering the correlation information that exists among the different metrics, then employ multiple-layer neural networks to study the characteristics of the typical samples. Finally, we justify the untrustworthy Web services through the designed bayesian-based probabilistic neural network.


\subsection{Neural Network}
\label{sec:4.3}

Artificial neural network is a common technique used in data analysis. Actually, in the service computing field, the neural network technique is employed to predict, discover, and classify Web services by combining with their QoS. Gao et al. \cite{g23} extended existing QoS model by adding new attributes that reflect performance of services and rely on artificial neural network (ANN) to provide client dynamic and on demand service performance prediction. Ahmed et al. \cite{g24} proposed a QoS-based model of ANN for Web services discovery, which combines an ANN based intelligent approach for publishing the QoS information and managing the reputation of Web services from customer feedback of their performance. Zhang et al. \cite{g21} proposed a global QoS-driven evaluation method based on artificial neural networks, aiming at facilitating the web service composition without preference weights. AI-Masri et al. \cite{g22} proposed a framework for enabling the efficient discovery of Web services to utilize well-known artificial neural networks (ANN) for their generalization capabilities. Through the aggregation of QoS of Web services, the neural network is capable of identifying those services that belong to a variety of QoS levels.

Based on these researches, we proposed a two-phase neural network model by combining QoS for classification and identification of untrustworthy Web services. The approach is different from \cite{g22}, which is based on the Web service relevancy function (WsRF) \cite{g25}. Our identification approach of untrustworthy Web services employs PNN by analyzing correlations among the QoS metrics and justifying trustworthy Web services by learning from the samples.

\section{Conclusion}
\label{sec5}

In this paper, we proposed a novel two-phase neural network model to identify the untrustworthy Web services. In the process, a feedforward neural network works as the classifier and a PNN works as the identifier to deciding untrustworthy Web services according to their QoS information. The experimental results indicated that the proposed approach has high accuracy compared with other competing approaches.

The study has successfully demonstrated that artificial neural network can be employed to detect untrustworthy Web services. However, there is room for improvement. It is observed that the feedforward neural network and PNN model are more time-consuming during the training mode in a large-size sample which could be an issue when implementing such an approach in real-time environments. Hence, in the future work, for further optimization purpose, we will consider to employ mini-batch training sample for decreasing the training time. In addition, we will explore other possible types of neural networks such as (Fuzzy Neural Network, FNN, and Convolutional Neural Network, CNN \etc.) for further enhancing the proposed approach.

\ifCLASSOPTIONcompsoc
\section*{Acknowledgement}
\else
\section*{Acknowledgment}
This work was partially supported by National Natural Science Foundation of China (No. 61272353, 61428201) and National Science Foundation-Career of USA (No. 1622292).

\ifCLASSOPTIONcaptionsoff
  \newpage
\fi


\begin{spacing}{0.85}
\bibliographystyle{IEEEtran}
\bibliography{Reference}
\end{spacing}

\end{document}